# Experimental and numerical study of the propagation of focused wave groups in the nearshore zone

Iskander Abroug [1,2], Nizar Abcha [2*], Denys Dutykh [3], Armelle Jarno[1], and François Marin[1]

[1]*Normandie Université, UNILEHAVRE, CNRS, UMR 6294 LOMC, 76600 Le Havre, France*
[2] *Normandie Université, UNICAEN, UNIROUEN, CNRS, UMR 6143 M2C, 14000 Caen, France*
[3] *Université Savoie Mont Blanc, CNRS UMR 5127, LAMA, 73000 Chambéry, France*

The propagation of focused wave groups in intermediate water depth and the shoaling zone is experimentally and numerically considered in this paper. The experiments are carried out in a two-dimensional wave flume and wave trains derived from Pierson-Moskowitz and JONSWAP spectrum are generated. The peak frequency does not change during the wave train propagation for Pierson-Moskowitz waves; however, a downshift of this peak is observed for JONSWAP waves. An energy partitioning is performed in order to track the spatial evolution of energy. Four energy regions are defined for each spectrum type. A nonlinear energy transfer between different spectral regions as the wave train propagates is demonstrated and quantified. Numerical simulations are conducted using a modified Boussinesq model for long waves in shallow waters of varying depth. Experimental results are in satisfactory agreement with numerical predictions, especially in the case of wave trains derived from JONSWAP spectrum.

1. Introduction

The study of the propagation of focused wave groups in the nearshore zone is a challenging task. Dispersive focusing is defined by the fast spreading of long waves and slow propagation of short waves [1, see *section 4.5*]. Several experimental studies have analyzed the evolution of wave groups in two-dimensional wave flumes using energy focusing in deep water [2-10]. This subject has also been numerically studied, in particular to analyse the breaking processes [9,11]. However, the propagation of dispersive focusing waves has been less investigated in intermediate water depth and coastal zones.

The estimation of energy transfers among different frequency components during the wave train propagation in the nearshore zone remains poorly understood. Rapp and Melville (1990) [2] were among the first to study the spatial evolution of wave frequency spectra. They found that the energy loss due to breaking is mainly from the high frequency end of the first harmonic region ($f_p < f < 2f_p$, where $f_p$ is the peak frequency). Baldock et al. (1996) [4] made similar observations and concluded that the focusing process is strongly nonlinear close to the focusing and breaking points. Tian et *al.* (2011) [11] studied experimentally and numerically the spatial evolution of frequency spectra of unidirectional narrow-banded waves (breaking and non-breaking) in deep water. A constant wave steepness spectrum (CWS) was adopted in their experimental and numerical focused wave study. They partitioned the spectrum into four regions to quantify the corresponding spectral variations. Nonlinear energy transfers and reversibility were found to occur between the transfer region ($1,2f_p < f < 1,5f_p$) and the higher frequency regions ($1,5f_p < f < 2,5f_p$) during the wave train propagation. A weak energy dissipation in the peak region ($0,9f_p < f < 1,1f_p$) and a slight energy gain in lower frequency region ($0,5f_p < f < 0,9f_p$) were experimentally demonstrated. When tracking the amount of energy contained in the low frequency components upstream and downstream the breaking, no obvious energy gain was observed in the lower frequency region. Thus, it was concluded that the energy gain in this region is explained by a possible nonlinear energy transfer and the breaking process may not necessarily increase the energy. Meza et al. (2000) [12] performed similar observations and concluded that there were no spectral changes in the peak region after the wave breaking.

Breaking waves may be induced by the propagation of wave trains of Benjamin and Feir (1967) [13] type, also known as modulational instability [8,14,15]. Another important mechanism of wave breaking is dispersive focusing, described by Longuet-Higgins (1974) [16], Rapp and Melville (1990) [2], and Drazen et al. (2006) [17]. This mechanism generates a compact wave group. The energy dissipation rate due to wave breaking has been considered by Melville (1994) [18], Drazen et al. (2008) [8] and Tian et al. (2010) [9].

Most of the aforementioned studies were conducted for deep water conditions, with wave trains often derived from unrealistic spectrum shapes, such as constant wave amplitude (CWA) [2,4,19], constant wave steepness (CWS) [19] and Gaussian distribution [20]. To the authors' knowledge, neither study has attempted to quantify the spectral change resulting from the propagation of realistic spectrum wave trains in the nearshore zone. It is then important to analyze the characteristics of focused waves generated by the focusing energy technique in intermediate and shallow water depth, and the propagation of such waves in the shoaling and breaking zones.

In the present study, detailed experiments are performed on numerous wave trains in order to determine the spatial evolution of peak frequencies related to waves propagating from a constant intermediate water depth to the shoaling and breaking zones, to study the spatial evolution of frequency spectra and to quantify the energy transfers between different frequencies. We also apply a simple hybrid Boussinesq-shallow water model [21] in order to compare the experimental results to the numerical predictions. Compared to the Saint-Venant equations, Boussinesq equations are derived under the assumption of non-hydrostatic pressure, resulting in a dispersive system of equations. In the dispersive focusing technique, the wave speed depends on the wavelength. For this reason, Boussinesq equations capture more physical effects than the classical shallow water equations [22].

The experimental set-up and the test conditions are presented in section 2. The spatial evolution of the peak frequency and frequency spectra are discussed in section 3. Numerical simulations and discussion are provided in section 4. Section 5 is devoted to conclusions and perspectives.

2. **Experimental setup and wave train parameters**

**Experimental setup**

The experiments were carried out in the wave flume of the M2C (Morphodynamique Continentale et côtière) Laboratory at Caen University, France. Focused wave groups are generated with an Edinburgh Designs Ltd piston type paddle using a linear wave generation signal. The wave flume is 22m long, 0.8m wide, 0.8m high and filled to a still water depth, $h$, of 0.3 m. The transformation of wave trains during their propagation towards the shore has been studied using an artificial slope $\beta=4\%$, which is placed 9.5 m downstream of the wave paddle in order to ensure intermediate water depth conditions before reaching the toe of the slope (Fig. 1).

The temporal variation of the free surface elevation is measured with wave gauges composed of two 50 cm long copper wires. Static calibrations made before and after each experiment are used to convert the measured voltage to surface elevation (in m). Two wave gauges are used in the experiments with a sampling rate of 50Hz. The first one (WG1), used as a temporal reference, is fixed at x = 4 m from the wave maker and the second one (WG2) is moved 20 cm downstream after each experiment (Fig. 1). The last measurement is situated at x =14 m from the wave paddle. This ensures a precise overview of the spatial evolution of a single wave train from the 50 measurements obtained along a distance of 10m (Fig. 1).

In these experiments, a finite length wave train of a duration $\Delta t$ =35s is generated. This time is sufficient because all meaningful wave components are taken into account. The used focusing technique is capable of producing wave trains with reasonable spatial and temporal repeatability. In fact, a calm water surface before the beginning of each test is necessary to achieve good repeatability. Prior to wave generation, the water level variation is monitored and limited to less than ±1.5mm. Therefore, at least five minutes are required between experiments to obtain a near quiescent state which is also inspected visually before each experiment. In order to reduce the influence of reflection when generating single wave trains, measurements are made before reflected waves travel back to the measurement point. In other words, slope do not affect the present results. Video tests are also performed in order to track the breaking process. The flume walls are made of transparent glass enabling wave profiles to be viewed and recorded using an external camera, capturing a field of view of 640x480 pixels at 100 frames/s. The spatial resolution is 0.467 mm/pixel. Observations with a high-speed camera show that the horizontal locations of wave breaking onset varies within ±1cm.

The linear NewWave profile [23] is used as input for the present focused wave groups. A wave train is composed of n sinusoidal wave components that come into phase at a single point in time and space. Using linear wave theory (LWT), the free surface elevation can be written as follows:

$$\eta(x,t) = \sum_{i=1}^{n} a_i \cos(k_i x - \omega_i t + \varphi_i). \tag{1}$$

where x is the distance to the mean wavemaker position, t the time, $a_i$ the wave amplitude of the $i^{th}$ component, $k_i$ the wave number, $\omega_i$ the wave angular frequency and $\varphi_i$ the phase angle at the focusing point. The input parameters are the wave spectrum type, the steepness, the phase at focusing ($\varphi=\pi$ or $\varphi=0$), the peak frequency ($f_p$) and the focus location. The generated wave trains are derived from a Pierson-Moskowitz (Eq. (2)) [24] or a JONSWAP (Eq. (3)) spectrum [25].

$$E_{Pierson-Moskowitz}(f) = \alpha g^2 (2\pi)^{-4} f^{-5} \exp\left(-\frac{5}{4}\left(\frac{f}{f_p}\right)^{-4}\right) \quad (2)$$

$$E_{JONSWAP}(f) = \alpha g^2 (2\pi)^{-4} f^{-5} \exp\left(-\frac{5}{4}\left(\frac{f}{f_p}\right)^{-4}\right)\gamma^\delta \quad (3)$$

where $\alpha = 0.0081$ corresponds to Phillips constant, $g$ is the gravity acceleration and $f_p$ denotes the peak frequency. The structure of the source functions representing the Pierson-Moskowitz and JONSWAP spectra are derived using field experiments data. The formula describing a JONSWAP spectrum (Eq. (3)) is obtained by multiplying a Pierson-Moskowitz spectrum with the peak-enhancement factor:

$$\gamma^\delta = \gamma^{\exp\left(-\frac{(f-f_p)^2}{2\sigma^2 f_p^2}\right)} \quad (4)$$

$$\sigma = \begin{cases} 0.07 \text{ for } f \leq f_p \\ 0.09 \text{ for } f \geq f_p \end{cases} \quad (5)$$

Where $\gamma$ is the ratio of the maximum spectral energy to the maximum of the corresponding Pierson-Moskowitz spectrum and $\sigma$ defines the left and the right sided widths of the spectrum [25]. Two different peak-enhancement factor were investigated during the experiments, i.e. $\gamma = 3.3, 7$. The lower value $\gamma = 3.3$ corresponds to the standard JONSWAP formulation and $\gamma = 7$ provides a narrower spectrum.

The strength of wave breaking is varied by changing the input wave steepness. Depending on the steepness, the breaking is located between x = 10.5 m and x = 14.3 m. Concerning the phase at focusing, a crest-focused wave group is obtained for $\varphi_i=0$, whereas a trough-focused wave group is obtained when $\varphi_i=\pi$ [26,27]. The theoretical focus location is fixed at x = 12 m to ensure focusing occurred on the slope. Seventeen wave trains, each measured at 50 locations along the flume, are generated varying the spectrum type, the steepness and the phase at focusing.

Let us consider the wave parameters associated with the wave groups: the local wave steepness ($S$), the characteristic wavenumber ($K_s$) and the characteristic frequency ($f_s$). The initial local wave steepness ($S_0$) used to characterize wave train nonlinearity, is determined from surface elevation measured at the first wave station (x = 4 m). Several definitions are given in previous studies to calculate this dimensionless parameter [8,18]. In the present work, the definition given in [9] is adopted:

$$S_0 = k_{s0} \sum_{i=1}^{n} a_i, \qquad (6)$$

where $\sum_{i=1}^{n} a_i$ is the surface elevation at the focusing point according to LWT and $k_{s0}$ is the spectrally weighted wavenumber related to the spectrally weighted frequency ($f_{s0}$) calculated at x = 4 m. This frequency is obtained as follows [9]:

$$f_{s0} = \frac{\sum_{i=1}^{n}[f_i a_i^2 (\Delta f)_i]}{\sum_{i=1}^{n}[a_i^2 (\Delta f)_i]}. \qquad (7)$$

where $f_i = \omega_i/2\pi$ and $\Delta f$ is the difference between components, which is constant ($\Delta f$ =0.0286). Based on linear wave theory, $f_i$, $a_i$ and $\Delta f$, are obtained by spectrum analysis performed on the free surface elevation measured along the flume through FFT (Fast Fourier Transform) using Eq. (8). Each measurement is decomposed into 70 Fourier components and the corresponding surface elevation is expressed analytically as a summation of sinusoidal waves in order to determine the different amplitudes ($a_i$) and frequencies ($f_i$) [9]. Finally, the linear finite water depth dispersion relation given by Eq. (9) is applied to obtain $k_{s0}$.

$$F(f) = \int_0^T \eta(t) e^{-2\pi i f t} dt. \qquad (8)$$

$$\omega_{s0}^2 = g k_{s0} \tanh(k_{s0} h), \; \omega_{s0}=2\pi f_{s0}. \qquad (9)$$

Fourier frequency components in the range [0.2, $3f_p$] Hz are included in the computation of $S_0$, $k_{s0}$ and $f_{s0}$. The latter parameters remain the same even when higher Fourier components are included. The different wave trains generated in this study are provided in Table. 1. The variable parameters are the spectrum type, the peak frequency $f_p$, the wave steepness and the phase at focusing (φ). The wave trains start propagating in intermediate water depth (0.75<$k_p$h<0.96, where $k_p$ is the wavenumber related to $f_p$).

The wave group is generated with a given linear focus position based on linear focusing in a constant water depth. It should be noted that the breaking on the slope prevents complete focusing. Subsequently, the focusing is not experimentally perfect. For most of the generated wave trains, more than one breaker occurs. The wave train breaks in the vicinity of the focusing and so breaking locations ($x_b$) will be indicated (Table 1).

The wave frequency spectrum in a given location is calculated as $S(f) = 2|F(f)|^2$. Fig. 2 exhibits three typical sets of the spatial evolution of wave spectra for Pierson-Moskowitz, JONSWAP (γ=3.3) and JONSWAP (γ=7) wave trains. The reference spectrum calculated at x = 4 m is plotted as a dotted line. A first observation to be made is that, as expected, the energy distribution varies from one spectrum type to another. For JONSWAP (γ=7) spectrum, the energy concentrates mainly near the peak frequency. However, for Pierson-Moskowitz, the frequency spectrum for $f>f_p$ decreases gradually as the frequency increases. Qualitatively, Fig. 2 shows a considerable change in the spectrum shape during the wave train propagation, upstream and downstream breaking, on the flat and on the sloping bottom. This change concerns all parts of spectra: the peak frequency ($f=f_p$), the high frequencies ($f>f_p$), the low frequency regions ($f<f_p$) and differs from one spectrum type to another. In addition, a slight downshift of the spectral peak frequency in the case of the two JONSWAP (γ=3.3 and γ=7)) wave spectra is observed. Quantitative results are given in section 3.

**Spectra partitioning into frequency regions**

In order to track spectral changes, Tian et al. (2011) [11] defined four frequency regions in the case of constant wave steepness (CWS) wave trains. This partitioning, simply based on spectral changes in a given location, is the following:

*Constant wave steepness [11]*

$$\begin{cases} \text{Peak region} & : \quad 0.9f_p < E1 < 1.1f_p \\ \text{Transfer region} & : \quad 1.1f_p < E2 < 1.5f_p \\ \text{High frequency region} & : \quad 1.5f_p < E3 < 2.5f_p \\ \text{Low frequency region} & : \quad 0.5f_p < E4 < 0.9f_p \end{cases}$$

For present tests, a partitioning is performed for each spectrum type at x = 4 m. E1, E2, E3 and E4 are respectively peak, transfer, high frequency and low frequency regions. Fig. 3(a) illustrates the four frequency regions obtained for a Pierson-Moskowitz wave train (PMW4). Fig. 3(b) shows the PMW4 spectrogram representing the spatial evolution of frequency spectrum from x = 4 m to x = 14 m

The present partitioning is based on the maximum frequency spectrum situated in the peak region (E1) at x = 4 m. The lower limit for the low frequency region (E4) is 0.2Hz and the upper limit corresponds to the lower limit of the peak region, which is set to be 20% of the maximum frequency spectrum. The selected lower and upper limits for E1 guarantee consideration of all significant frequency spectrum around the peak frequency. The transition to the transfer region (E2) is marked by a significant decrease in frequency spectrum (i.e. 60% of the maximum frequency spectrum). The upper limit of E2 coincides with 20% of the maximum frequency spectrum. Subsequently, the high frequency region (E3) is situated beyond the transfer region and the upper limit of E3 is set to $3f_p$. The frequency spectrum situated beyond this region is very small (less than 2% of the total frequency spectrum) and can be neglected. The frequency limits for the E1 to E4 regions are given below for the Pierson-Moskowitz and JONSWAP ($\gamma=3.3$ and $\gamma=7$) spectra. In Fig. 4, a partitioning is applied to spectrograms of four wave trains with the following limits according to the spectrum type:

**Pierson-Moskowitz spectrum**

| | | |
|---|---|---|
| Peak region | : | $0.7f_p < E1 < 1.25f_p$ |
| Transfer region | : | $1.25f_p < E2 < 1.85f_p$ |
| High frequency region | : | $1.85f_p < E3 < 3f_p$ |
| Low frequency region | : | $0.2 < E4 < 0.7f_p$ |

**JONSWAP spectrum ($\gamma=3.3$)**

| | | |
|---|---|---|
| Peak region | : | $0.7f_p < E1 < 1.1f_p$ |
| Transfer region | : | $1.1f_p < E2 < 1.45f_p$ |
| High frequency region | : | $1.45f_p < E3 < 3f_p$ |
| Low frequency region | : | $0.2 < E4 < 0.7f_p$ |

**JONSWAP spectrum ($\gamma=7$):**

| | | |
|---|---|---|
| Peak region | : | $0.75f_p < E1 < 1.05f_p$ |
| Transfer region | : | $1.05f_p < E2 < 1.25f_p$ |
| High frequency region | : | $1.25f_p < E3 < 3f_p$ |
| Low frequency region | : | $0.2 < E4 < 0.75f_p$ |

3. **Experimental results**

**The spatial evolution of the peak frequency**

In this section, the variation of the peak frequency during the propagation is discussed for different spectra and for different levels of steepness ($S_0$). In order to carefully monitor its spatial evolution, $f_p$ is normalized by $f_{p0}$ calculated at x = 4 m. Fig. 5 exhibits the spatial variation of the peak frequency for the studied wave trains.

Recently, Liang et *al.* (2017) [19] studied the spatial evolution of the peak frequency for Pierson-Moskowitz and constant wave steepness spectrum in constant deep water. They found that $f_p$ is constant along the wave train propagation and not affected by the breaking intensity, the frequency width, or spectral type. Fig. 5(a-b) show the variation in normalized peak frequency with distance for different Pierson-Moskowitz wave trains. The experimental peak frequency remains nearly stable upstream and in the breaking zone (4 m < x < 14 m) for all tested Pierson-Moskowitz wave trains. Similar conclusions were made by Stansberg (1994) [28] who studied broad-banded and narrow-banded bichromatic deep-water waves. This behaviour is the same on the flat and the sloping bottom. However, a slight downshift (5%) is observed after breaking at x = 14 m for group focused waves with trough or crest at focusing. Hence, the phase at focusing does not impact the peak frequency evolution for Pierson-Moskowitz wave trains.

Peak frequency variations in the case of JONSWAP ($\gamma$=3.3 and $\gamma$=7) wave trains are also quantified and exhibit a decreasing rate which depends on the steepness ($S_0$). During its propagation, the wave train naturally tends to a more stable configuration via a downshift of the energy. In other words, the decrease is directly related to energy transfer from the higher to lower frequency ranges. For the two steeper JONSWAP wave trains, $f_p$ decreases along the flat and the sloping bottom and reaches 85% of its initial level for JSW3 and 81% for JSW6 at x = 14 m**.** To the best of our knowledge, similar results have not been reported in previous studies.

In the literature, it is commonly agreed that the spectral peak downshift is caused by a combination of nonlinear wave modulation, dissipation and breaking. The spectrum energy tends to shift to a lower frequency as the wave train propagates along the flume due to the sideband instability.

**The spatial evolution of the frequency spectrum**

For a given wave train, measurements begin and end in quiescent conditions. Consequently, the time integration of the free surface elevation ($\Delta t$=35s) provides the total energy. In other words, all the non-zero surface elevations of the generated wave train ($\eta(x,t) \neq 0$) are included to calculate the frequency spectrum. In this section, all energy forms are normalized by $E0_1$, which is the integral of the wave frequency spectrum $S(f)$ between the cut-off frequencies [0.2, $3f_p$] Hz and measured at the first wave station (x = 4m). Energy variations are quantified as a function of space every 20 cm from x = 4m up to x = 14 m. The spatial variation of the energy in E1, E2, E3 and the total energy E0 are illustrated in Fig. 6, Fig. 7 and Fig. 8. The two vertical solid lines indicate the breaking region.

The first observation to be made is that the total energy E0 decreases gradually on the flat bottom (4 m < x < 9.5 m) for all studied wave trains and between 5% and 15% of E0 is dissipated. This energy loss in the flat bottom is mainly due to surface damping and friction by the flume sidewalls and bottom. After the breaking, where x > $x_b$, the stronger the breaking, the greater the total energy dissipation at the end of the wave propagation (x = 14 m). For instance, the loss ratio is about 15% of the total energy for the weakest Pierson-Moskowitz wave train (P-MW1) and 70% in the case of the strongest one (P-MW4). Similarly, the loss ratio is around 60% of the total energy for JSW33 and 35% for JSW22.

The initial energy amount in E1 is between 50% and 70% of the total energy (E0). Over the flat bed (4 m < x < 9.5 m), the spectral peak energy of Pierson-Moskowitz wave trains increases slightly and reaches 1.2 times its initial level for P-MW44 at the toe of the beach. This could be a way of compensating for energy dissipation in the transfer region (E2). Liang et al. (2017) [19] made qualitatively similar observations in the case of three types of deep-water wave spectra (i.e. constant-amplitude, constant-steepness, and Pierson-Moskowitz). The spectral peak energy (E1) remains approximately constant in the case of JONSWAP wave trains along the flat bottom.

A rapid dissipation introduced by wave breaking is illustrated in Fig. 6, Fig. 7 and Fig. 8. These results suggest that wave breaking has an immediate impact on the spectral peak region (E1). The decrease rate is directly proportional to the wave train steepness ($S_0$). For example, the loss ratio is about 70% of its initial level for P-MW4 and around 50% for the JONSWAP wave train with similar nonlinearity (JSW55).

The spatial evolution of energy in the transfer region (E2) and in the high frequency region (E3) is also investigated. The initial energy amount is between 20% and 35% of the total energy (E0) in E2 and between 10% and 20% in E3, which is in good agreement with [11]. It is found that the energy in E2 decreases gradually and significantly upstream of breaking (4 m < x < $x_b$). This behaviour is qualitatively similar for all steepness and for both studied spectra (Fig 6, Fig. 7 and Fig. 8). No noticeable change in the decreasing trend is observed during the wave shoaling (9.5 m < x < $x_b$). Just prior to the breaking, the amount of energy in E2 reaches its minimum level. Losses found in the transfer region are partially due to surface damping and friction by the flume sidewalls and bottom. Moreover, energy contained in E2 spreads to E3 as the wave train evolves to the focal point. The observed energy transfer from the first harmonic to higher harmonics is consistent with earlier studies [4,11,29,30,31]. Hence, the spatial behaviour of energy in E2 and E3 in intermediate and shallow water depths is qualitatively similar to that in deep water depth. Quantitatively, from 2 to 9 % of the total energy is transferred from E2 to E3 depending on the breaking strength. Downstream of breaking (x > $x_b$), the energy contained in E2 starts to increase in the case of JONSWAP ($\gamma$=7 and $\gamma$ =3.3) wave trains, which is clearly noticeable for the strongest steepness (JSW3, JSW33 and JSW6). This finding can be explained by the reversibility of the nonlinear energy transfer between E2 and E1. Concerning E3, the energy contained in this region tends to zero, which is quantitatively similar to observations made in [11].

Tian et al. (2011) [11] observed energy gain in low frequency components during wave train propagation. They suggested that this is possibly due to nonlinear energy transfer from the transfer region to the lower frequency region. Longuet-Higgins (1969) [32], Hasselmann (1971) [33] and Melville (1996) [34] attempted analytically to understand how energy is transferred from high to low frequency components during the breaking process. According to [34], this energy transfer might be due to the release of free long wave components arising from breaking. Meza et al. (2000) [12] were the first to attempt to quantify experimentally the energy transferred to low frequency components during the breaking process. They found that the wave components $f < f_p$ gain around 12% of the energy lost by the wave components $f > f_p$. Yao and Wu (2004) reached similar conclusions by simply comparing measurements upstream and downstream of breaking in deep water depth and for JONSWAP wave trains.
Fig. 9 indicates that the energy in the region E4 appears to increase during the propagation (4 m < x < $x_b$) for most of the breaking wave trains, which is consistent with the results obtained by other authors [2,11,12]. This increasing trend is due to a nonlinear energy transfer from higher frequency regions (E1 and E2) to the lower frequency region (E4). With the breaking intensity strengthening, energy contained in E4 decreases slightly when the wave train breaks. This is clearly noticeable for the steepest JONSWAP wave trains (JSW3, JSW33, and JSW6). This could be a way of compensating energy dissipation in the peak region.

To conclude, the spectral gain in E4 varies slightly from one spectrum to another. At the end of the propagation (x = 14 m), the wave components below the peak region increase by a maximum of 12 % of the total energy for the Pierson-Moskowitz spectrum, 10% for the JONSWAP spectrum with γ=3.3 and 8% for the JONSWAP spectrum with γ=7.

4. **Numerical simulations and discussion**

An enhanced version of the model developed by Dutykh et al. (2011) [35] is used herein to reproduce the laboratory experiments in order to provide further insight into the measured wave behavior. Dutykh et al.'s model solves an extended form of the classical Boussinesq equations derived by Peregrine [36], with higher order nonlinear terms used to ensure correct linear dispersion. It is assumed the fluid is incompressible and the flow irrotational. The equations are modified so that the governing equations are invariant (by vertical translation) with respect to choice of still water level. Full details of the shock-capturing adaptive Runge-Kutta finite volume solver, called mPeregrine, and its validation against experimental run-up data are given in [21] and [35] respectively.

Under the previously described physical assumptions, Dutykh et al. (2011) [35] derived the following system of equations based on Peregrine's system:

$$\begin{cases} H_t + Q_x = 0 \\ Q_t + (\frac{Q^2}{H} + \frac{g}{2} H^2)_x - P(H,Q) = gHh_x \\ P(H,Q) = \frac{H^2}{3} Q_{xxt} - \frac{1}{3}\left(H_x^2 - \frac{1}{6} HH_{xx}\right) Q_t + \frac{1}{3} HH_x Q_{xt} \end{cases} \quad (10)$$

where $H(x,t) = \eta(x,t) + h(x)$ is the total water depth, $h(x)$ represents the depth below the mean sea level, $Q(x,t) = H(x,t)u(x,t)$ is the horizontal momentum, $u(x,t)$ is the depth-averaged fluid velocity and under-scripts ($H_t \stackrel{\text{def}}{=} \frac{\partial H}{\partial t}, Q_x \stackrel{\text{def}}{=} \frac{\partial Q}{\partial x}$) denote partial derivatives. We mention that ignoring the dispersive terms $P(H,Q)$ of system 10 we obtain the classical shallow water equations.

In the present simulations we use 2000 cells, a time step of 0.2 s, and carry out computations on a domain length 20 m for a total simulation time of 35 s. Initial conditions are prescribed using experimental free surface signals obtained at x = 4 m. The model is calibrated by altering the friction coefficient until close agreement is obtained between predicted and measured wave date. This is performed by including the appropriate dissipative term $F(u,H) = -c_m g \frac{u|u|}{H^{1/3}}$ into momentum conservation equations (Eq. (10)). The calibrated value of Manning's coefficient is $c_m$ = 0.01 s.m$^{-1/3}$ for a still water depth of 0.3 m, which is close to the expected value of 0.009 s.m$^{-1/3}$ for plastic. Numerical simulations are carried out in order to evaluate the accuracy of the mPeregrine model with respect to spatial variations in peak frequency and spectral energy. It should be noted that less agreement is obtained for the frequency spectrum than the surface elevation time series because discrepancies between experimental and predicted free surface elevations are amplified because the spectrum relates to the squared wave amplitude.

Fig. 10 presents the experimental (black) and predicted (red) energy spectra at selected locations along the flume for Pierson-Moskowitz (left), JONSWAP ($\gamma$=3.3) (middle), and JONSWAP ($\gamma$=7) spectrum (right) spectra. Qualitatively, the model accurately predicts the peak frequency downshift and the spectrum computed on flat bottom, on sloping bottom, upstream and downstream of breaking in the case of JONSWAP ($\gamma$=7) spectrum. Nevertheless, discrepancies become larger for JONSWAP ($\gamma$=3.3) and Pierson-Moskowitz spectra and are enhanced close to the wave breaking.

Fig. 11 shows that the predicted maximum peak frequency downshift is in qualitative agreement with that observed experimentally (Fig. 5) for all spectra, with differences within 0-10%. It is evident from Fig. 11(b) that the numerical evolution of the peak frequency for the steepest wave train (JSW3) exhibits slumps, which are represented by vertical solid lines. The presence of these slumps in the spatial evolution of peak frequency of large amplitude wave trains could be related to the uncontrolled numerical instabilities arising from the discretization of dispersive terms near the shoreline [21,37]. To conclude, the mPeregrine code has accurately predicted the spatial variation of the peak frequency in the case of Pierson-Moskowitz and JONSWAP ($\gamma$=3 and $\gamma$=7) wave trains.

In order to evaluate the accuracy of the numerical code in predicting the spectral evolution of wave groups, the same partitioning performed in section II is adopted to track energy variations in different frequency ranges. Fig. 12 and Fig. 13 represent comparisons between experimental and numerical spectrograms.

Numerical predictions of energy in the peak (E1) and transfer regions (E2) are satisfactory for the case of JONSWAP ($\gamma$=7) wave trains (e.g. JSW5) and the difference is within 10-20% (Fig. 12). Yet, the discrepancies become large in the case of Pierson-Moskowitz wave trains

(Fig. 13). A possible explanation might be the large width of E1 ($0.7f_p < E1 < 1.25f_p$) in the Pierson-Moskowitz spectrum. The magnitude of the frequency spectrum in E3 is underestimated in numerical predictions for all wave steepness. This is expected because the equations underpinning the mPeregrine model are derived to represent localised, long, dispersive waves in shallow water. Consequently, the mPeregrine code reproduces satisfactorily the spatial evolution of energy in E2 and E1 as long as their width is narrow; the model is less robust for frequencies $f > 1.4f_p$.

Moreover, the increase in E4 is qualitatively tracked by numerical simulations. Nonetheless, this agreement is not quantitatively satisfactory for $f < 0.6f_p$ and important discrepancies are observed (within 50%) caused by spurious waves (i.e. sub-harmonic error) generated by the wave maker [27,38].

To conclude this part, this simple model with constant Manning's roughness gives satisfactory predictions for frequencies between $0.6f_p$ and $1.4f_p$. In other words, the agreement between numerical results and experiments remains adequate because E1 and E2 regions are narrow.

## 5. Conclusions and perspectives

This paper presents a study of the propagation of focused wave groups above an artificial bed before and after breaking using a series of physical experiments and numerical simulations. The influence of the steepness ($S_0$), the spectrum type and the phase at focusing ($\varphi$) on focused wave characteristics are examined.

Present results show that the peak frequency obtained experimentally remains stable during the wave train propagation for Pierson-Moskowitz wave trains, whereas an obvious downshift was observed in the case of JONSWAP ($\gamma=3.3$ and 7) wave trains. The downshift can be interpreted as a self-stabilizing feature of the process. Its decreasing rate is more pronounced when the wave steepness ($S_0$) is very high and the peak region (E1) is narrow. The most important modulations are observed in narrow-banded wave train spectra, while for broad-banded spectra, i.e. Pierson-Moskowitz, energy content seems to be more stable and the peak frequency remains approximately constant during the wave train propagation.

In order to track the energetic behaviour of the recorded wave trains in different frequency ranges, four energy regions are defined. Regions E1, E2, E3 and E4 are respectively the peak, transfer, high frequency and low frequency regions. Along the flat bottom (4 m < x < 9.5 m), between 5% and 15% of the total energy (E0) is dissipated at the toe of the beach (x = 9.5m) due to friction by the flume sidewalls and bottom. This energy loss is mainly from the transfer region E2. Of particular interest is the slight increase of energy in the peak region E1 in the case of Pierson-Moskowitz wave trains. This behaviour could be a way of compensating the strong energy dissipation in the transfer region. Energy is transferred to the surrounding upper and lower frequencies of $f_p$, which is consistent with the stable behaviour of the peak frequency in the case of Pierson-Moskowitz wave trains.

Along the wave train propagation (4 m < x < $x_b$), a nonlinear energy transfer from E2 to E3 is further observed. This is particularly noticeable when the wave steepness is very high. After breaking (x > $x_b$), a considerable change in spectrum shape occurs and a loss of energy in E3 is observed indicating an energy transfer to lower frequency regions (E2). The energy becomes concentrated in the narrow frequencies range near the peak frequency region (E1) and to a lesser degree near the transfer region (E2). These conclusions are consistent with experimental results found in deep water [11].

In addition, nonlinear energy transfer occurs from the peak region E1 to the low frequency region E4 during the wave train propagation (4 m < x < $x_b$). Between 8% and 12% of the total energy is transferred to E4 and no obvious decrease was found just after the breaking (x > $x_b$). These findings indicate a qualitative similarity in the energetic behaviour between the two tested spectra, and this confirms the defined partitioning introduced at the beginning of section 2.

Complementary numerical simulations based on a Boussinesq-type model for long waves are conducted in shallow waters of varying depth. The model was calibrated using linear generation signals by altering the friction coefficient, with the best agreement achieved at $c_m = 0.01$ s.m$^{-1/3}$. The numerical predictions are satisfactory in the prediction of peak frequency and spectral changes along the wave train propagation for frequencies between $0.6f_p$ and $1.4f_p$. Outside this frequency range, the wave frequency spectra have some discrepancies with the measurements. In other words, by narrowing the spectrum, the real wave shape is closer to the predicted one.

An inherent disadvantage of the dispersive focusing technique is the problem of scaling the findings to full scale. This kind of problem can be addressed by carrying out measurements at different scales. For the near future, we plan to perform new tests involving higher wave steepness, larger frequency bandwidths and to improve the numerical model in the case of random unidirectional sea in order to extend the simulations to wider spectra, in particular for the estimation of coastal responses to extreme waves.

## Acknowledgements

We thank the Normandy region for its financial support.

# References

[1] C. Kharif, E. Pelinovsky, A. Slunayev, Rogue waves in the ocean: Observations, Theories and Modelling, Springer, pp. 255 (2008), advances in geophysical and environmental mechanics and mathematics.

[2] R. J. Rapp, W. K. Melville, laboratory measurements of deep-water breaking waves, Philos. Trans. R. Soc. Lond. A 331 (1990), 735-800.

[3] C. Lin, D. Rockwell, Evolution of a quasi-steady breaking waves, J. Fluid Mech. 302(1995), 29-44.

[4] T. Baldock, T. E. Swan, P. H. Taylor, A laboratory study of non-linear surface waves on water, Philos. Trans. R. Soc. Lond. A 354(1996), 1-285.

[5] M. Perlin, J. He, L. P. Bernal, An experimental study of deep water plunging breakers, phys. Fluids 8(9) (1996). 2365-2374.

[6] H. M. Nepf, C. H. Wu, E. S. A. Chan, A comparison of two and three dimensional wave breaking, J. Phys. Oceanogr, 28(1998), 1496-1510.

[7] T. B. Johannessen, C. Swan, A laboratory study of the focusing of transient and directionally spread surface water waves, *in Proceedings of the Royal Society A, London, Ser, 457(2001), 971-1006.*

[8] D. A. Drazen, W. K. Melville, L. Lenain, Inertial scaling of dissipation in unsteady breaking waves, J. Fluid Mech. 611(2008), 307-332.

[9] Z. Tian, M. Perlin, W. Choi, Energy dissipation in two-dimensional unsteady plunging breakers and an eddy viscosity model, J. Fluid Mech, (2010), pp. 217-275.

[10] D. Liu, Y. Ma, G. Dong, M. Perlin, Detuning and wave breaking during nonlinear surface wave focusing, Ocean Eng, 113(2016), 215-223.

[11] Z. Tian, M. Perlin, W. Choi, Frequency spectra evolution of two-dimensional focusing wave groups in finite water depth water, J. Fluid Mech. 688(2011), 169-194.

[12] E. Meza, J. Zhang, R.J. Seymour, Free-wave energy dissipation in experimental breaking waves, J.Phys. Oceanogr. 30(2000), 2404-2418.

[13] T. B. Benjamin, J. E. Feir, Disintegration of wave trains on deep water, Part1, Theory. J. Fluid Mech. 27(1967), 27, 417-430.

[14] M. P. Tulin, T. Waseda, Laboratory observations of wave group evolution, including breaking effects, J. Fluid Mech, 378(1999), 197-232.

[15] M. L. Banner, W. L. Peirson, Wave breaking onset and strength for two-dimensional deep-water wave groups, J. Fluid Mech. 585(2007), 93-115.

[16] M. S. Longuet-Higgins, Breaking waves in deep or shallow water, in: Proc. 10[th] Conf. on Naval hydrodynamics, MIT, pp. 597-605 (1974).

[17] D. Drazen, Laboratory studies of nonlinear and breaking surface waves, University of California, San Diego, p. 202(2006).

[18] W. K. Melville, Energy-dissipation by breaking waves, J. Phys. Oceanogr. 24(1994), 2041-2049.

[19] S. Liang, Y. Zhang, Z. Sun, Y. Chang, Laboratory study on the evolution of waves parameters due to wave breaking in deep water, Wave Motion 68, 31-42 (2017).

[20] F. Günther, Claus, B. Jan. Gaussian wave packets – a new approach to seakeeping tests of ocean structures, Applied Ocean Research 8(4) (1986), 190-206.

[21] A. Duràn, D. Dutykh, D. Mitsotakis, Peregrine's System Revisited, N. Abcha, E. N. Pelinovsky, and I. Mutabazi (Eds.): Nonlinear Waves and Pattern Dynamics, pp. 3-43(2018), champ: Springer International Publishing.

[22] D. Dutykh, C. Dider. Modified shallow water equations for significantly varying seabeds, Applied mathematical modelling (40) (2016), 9767-9787.

[23] P. S. Tromans, A. R. Anaturk, P. Hagemeijer, A new model for the kinematics of large ocean waves-application as a design wave, Proceedings of the first international offshore and polar engineering conference, the international society of offshore and polar engineers, pp. 64-71 (1991).

[24] J. Jr. Pierson. Willard, L. Moskowitz, Proposed spectral form for fully developed wind seas based on the similarity theory of S.A. Kitaigorodskii, journal of geophysical research, Vol. 69, p. 5181-5190.

[25] K. Hasselmann, T. Barnett, E. Bouws, H. Carlson, D. Cartwright, K. Enke, J. Ewing, H. Gienapp, D. Hasselmann, P. Krusemann, A. Meerburg, A. Muller, P. Olbers, K. Richter, W. Sell, H. Walden. (1973). Measurements of wind-wave growth and swell decay during the Joint North Sea Wave Project (JONSWAP). Deutsches Hydrographisches Institut, Volume 12.

[26] A. C. Hunt-Raby, A. G. L. Borthwick, P. K. Stansby, P. H. Taylor, Experimental measurement of focused wave group and solitary wave overtopping, J. Hydraul. Res. 49(2011), 450-464.

[27] C. N. Whittaker, C. J. Fitzgerald, A. C. Raby, P. H. Taylor, J. Orszaghova, A. G. L. Borthwick, Optimisation of focused wave group runup on a plane beach, Coast. Eng. 121(2017), 44-55.

[28] C. T. Stansberg, Effects from directionality and spectral bandwidth on nonlinear spatial modulations of deep-water surface gravity wave trains, Coastal Engineering, Proceedings of the XXIV international conference, Kobe, Japan, 23-28, (1994), pp. 579-593.

[29] J. H. Kway, Yloh, E. Chan, Laboratory study of deep-water breaking waves, Ocean Eng. 25(8) (1998), 657-676.

[30] A. Yao, C. H. Wu, Energy dissipation of unsteady wave breaking on currents, J. Phys. Oceanogr. 34, 2288-2304 (2004).

[31] S. Liang, S. Zhaochena, Z. Yihui, S. Jiafa, Y. Zhang, Laboratory study on the characteristics of Deep-water Breaking Waves, *in Procedia Engineering, 116(2015), 414-421*.

[32] M. S. Longuet-Higgins, A nonlinear mechanism for the generation of sea waves, Proc. Roy. Soc. London, 311A(1969), 371-389.

[33] K. Hasselmann, On the mass and momentum transfer between short gravity waves and larger-scale motions, Journal. Fluid Mech. 50(1971), 189-205.

[34] W. K. Melville, The role of surface-wave breaking in air-sea interaction, Annu. Rev. Fluid. Mech, 28(1996), 279-321.

[35] D. Dutykh, T. Katsaounis, D. Mitsotakis, Finite volume schemes for dispersive wave propagation and runup, J. Comput. Phys. 230(8) (2011), 3035-3061.

[36] D. H. Peregrine, Long waves on a beach, J. Fluid Mech. 27(1967), 815-827.

[37] G. Bellotti, M. Brocchini, On the shoreline boundary conditions for Boussinesq-type models, Int. J. Numer. Methods Fluids 37(4) (2001): 479-500.

[38] J. Orszaghova, P. H. Taylor, A. G. L. Borthwick, A. C. Raby, Importance of second-order wave generation for focused wave group runup and overtopping, Coast. Eng. 94(2014), 63-79.

# Figure captions

1. Fig. 1. Side view of the flume. WG1 and WG2 are the two used wave gauges.
2. Fig. 2. The experimental dimensionless frequency spectrum for Pierson-Moskowitz, JONSWAP ($\gamma$=3.3) and JONSWAP ($\gamma$=7) spectrum. The reference spectrum calculated at x = 4 m is plotted as a dotted line. For the three wave trains, the wave breaking occurs between x = 12.07m and x = 12.81m.
3. Fig. 3. (a) The partitioning performed on the spectrum frequency relative to a Pierson-Moskowitz wave train (P-MW4) computed at x = 4 m. (b) The spectrogram relative to P-MW4. Concerning the spectrogram, the horizontal coordinate is the normalized frequency and the vertical coordinate is the abscissa (m) along the flume. The colorbar indicates the dimensionless wave frequency spectrum from x = 4 m to x = 14 m.
4. Fig. 4. Spectrograms obtained by wave gauges through FFT.
5. Fig. 5. The experimentally measured spatial peak frequency evolution of (a) Pierson-Moskowitz wave trains with crest at focusing, (b) Pierson-Moskowitz wave trains with trough at focusing, (c) JONSWAP ($\gamma$ =3.3) wave trains and (d) JONSWAP ($\gamma$ =7) wave trains.
6. Fig. 6. The spatial evolution of the dimensionless energy relative to Pierson-Moskowitz wave trains: squares correspond to the total energy E0, triangles represent the energy in the peak region E1, circles exhibits the energy in the transfer region E2, plus sign (+) illustrates the energy in high frequency region E3.
7. Fig. 7. The spatial evolution of the dimensionless energy relative to JONSWAP ($\gamma$=3.3) wave trains: squares correspond to the total energy E0, triangles represent the energy in the peak region E1, circles exhibits the energy in the transfer region E2, plus sign (+) illustrates the energy in high frequency region E3.
8. Fig. 8. The spatial evolution of the dimensionless energy relative to JONSWAP ($\gamma$=7) wave trains: squares correspond to the total energy E0, triangles represent the energy in the peak region E1, circles exhibits the energy in the transfer region E2, plus sign (+) illustrates the energy in high frequency region E3.
9. Fig. 9. The spatial evolution of the dimensionless energy in E4 for Pierson-Moskowitz and JONSWAP wave trains ($\gamma$=7 and $\gamma$ =3.3).
10. Fig. 10. The experimental (black) and the numerical (red) dimensionless frequency spectrum for Pierson-Moskowitz, JONSWAP ($\gamma$=3.3) and JONSWAP ($\gamma$=7) spectrum. For the three wave trains, the wave breaking occurs between x = 12.07 m and x = 12.81 m.

11. Fig. 11. The numerically predicted spatial peak frequency evolution of (a) Pierson-Moskowitz wave trains with crest at focusing, (b) JONSWAP ($\gamma$ =3.3) wave trains, (c) JONSWAP ($\gamma$ =7).
12. Fig. 12. Comparison of spectrograms obtained by wave gauges and those predicted by the mPeregrine code for JONSWAP wave trains ($\gamma$=3.3 and $\gamma$=7). The figures on the left-hand column present experimental results whereas numerical predictions are presented on the right-hand column.
13. Fig. 13. Comparison of spectrograms obtained by wave gauges and those predicted by the mPeregrine code for Pierson-Moskowitz wave trains. The figures on the left-hand column present experimental results whereas numerical predictions are presented on the right-hand column.

# Table captions

Table. 1. Wave train parameters

**Figures**

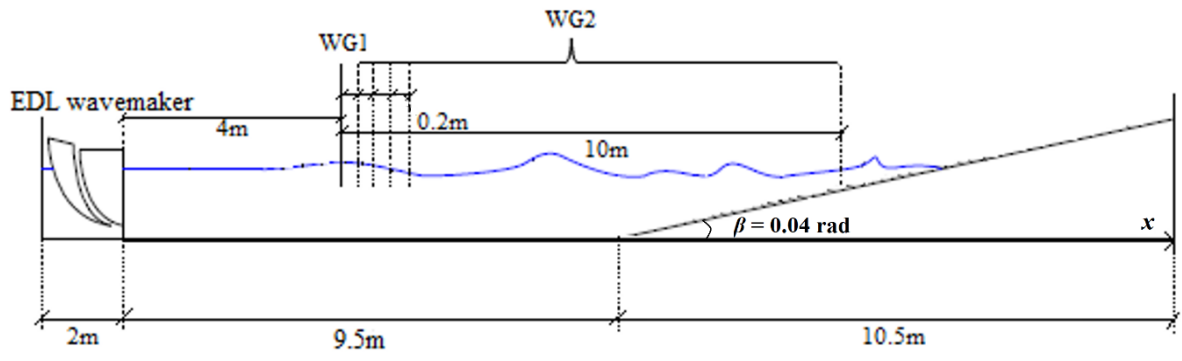

**Fig.1**

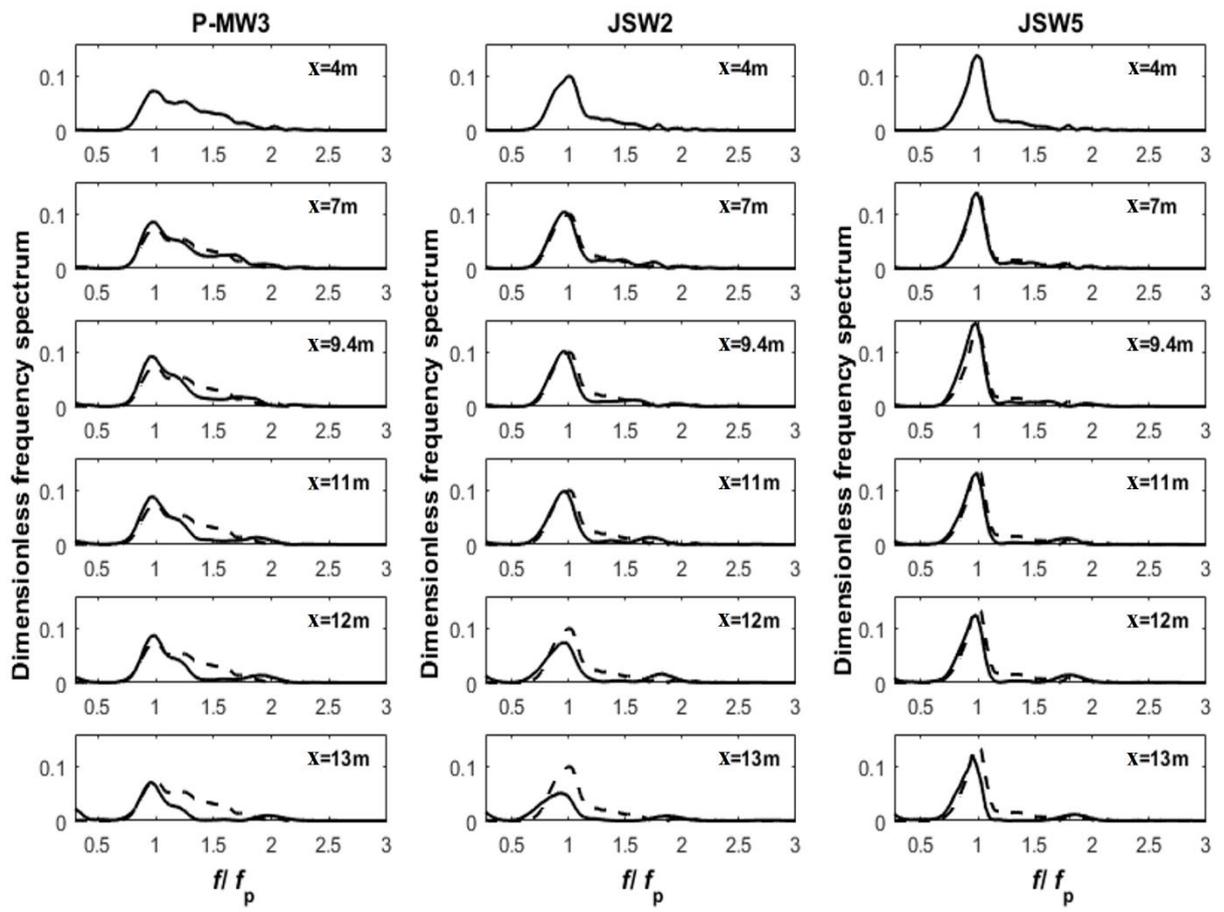

**Fig. 2**

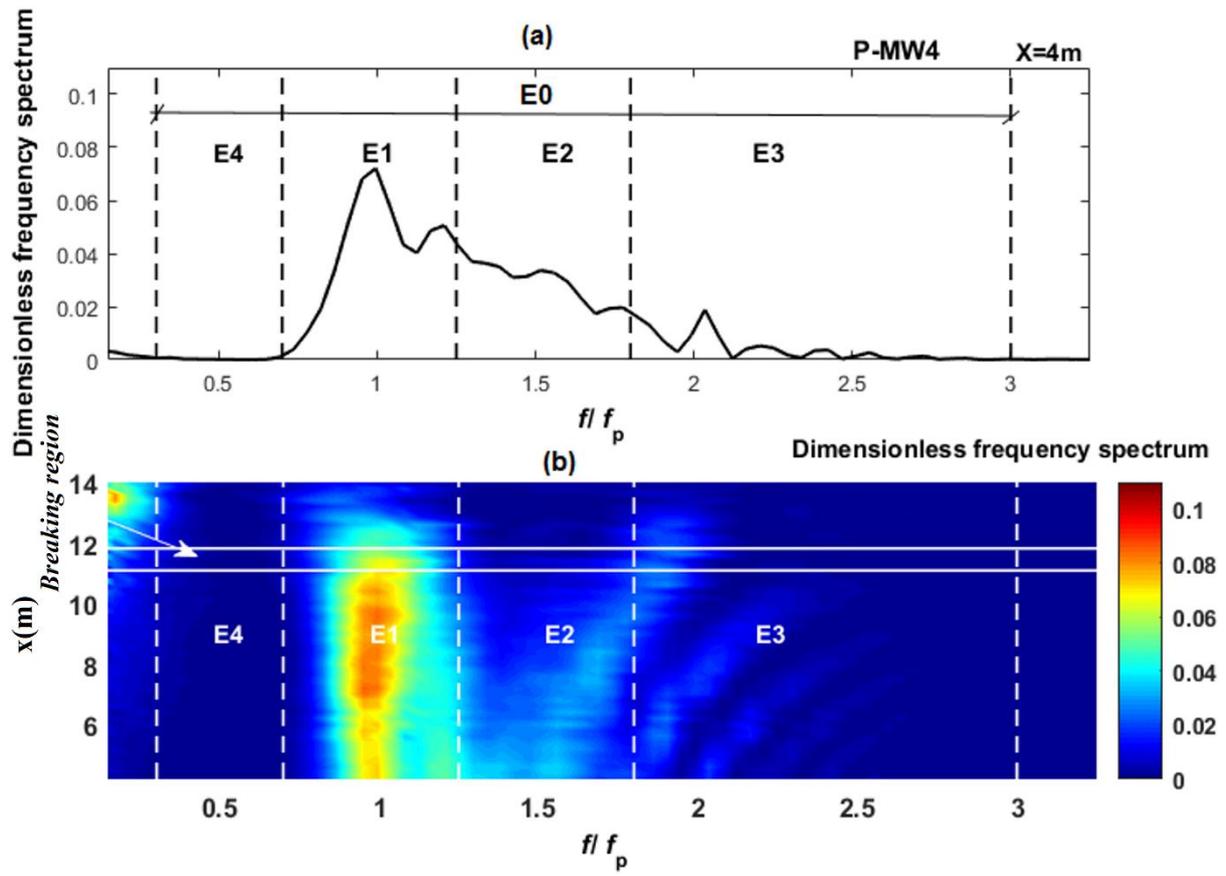

**Fig. 3**

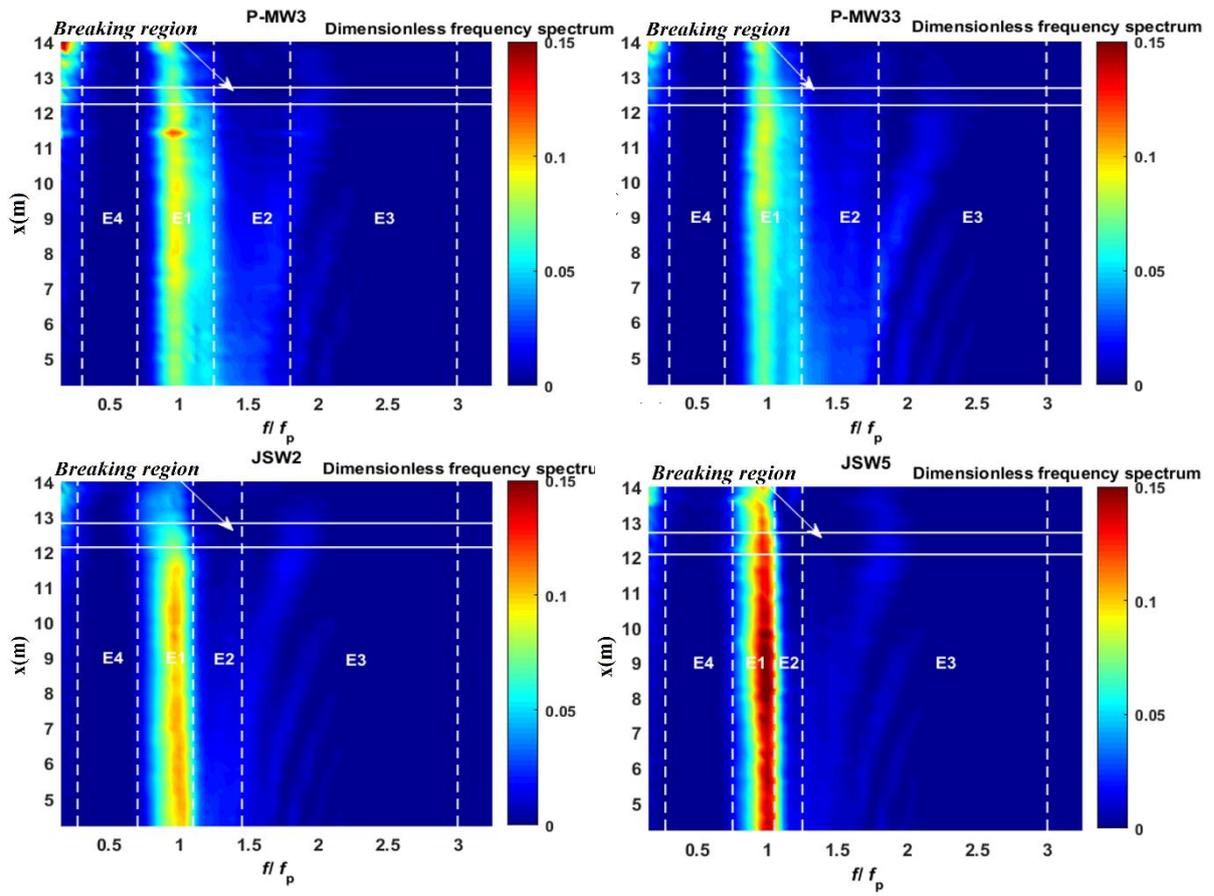

**Fig. 4**

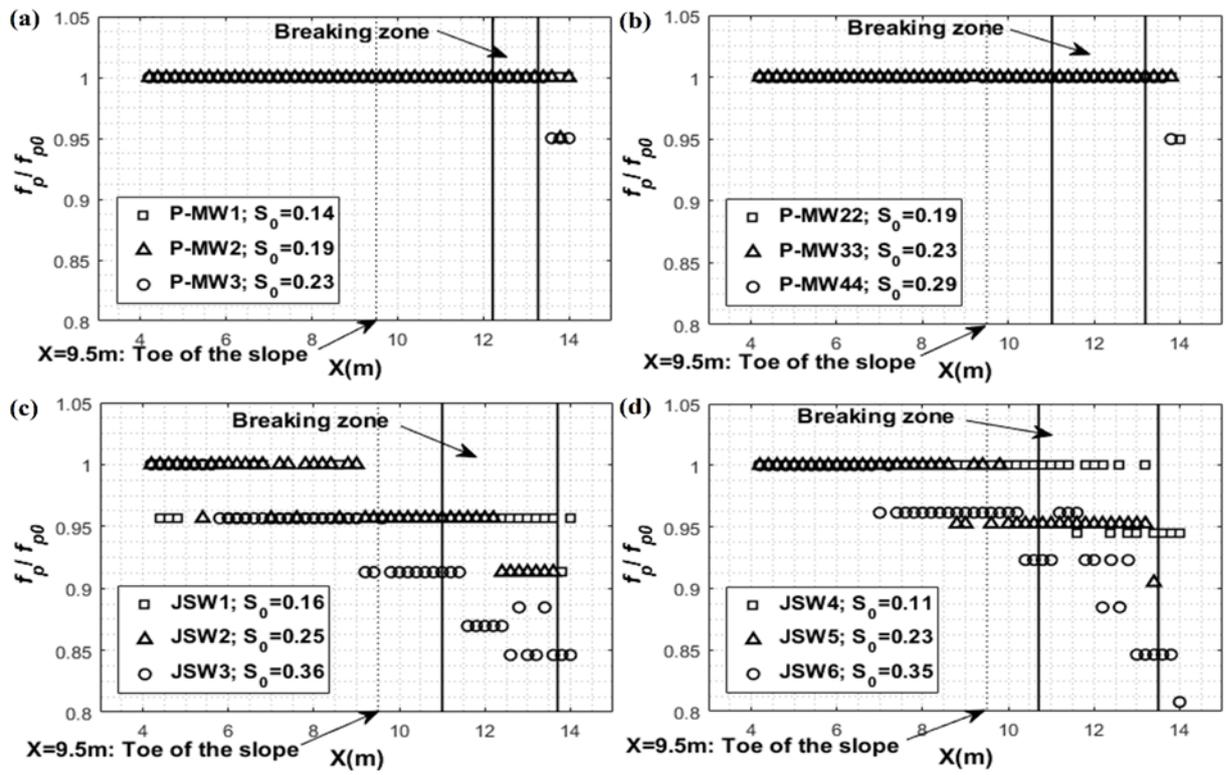

**Fig. 5**

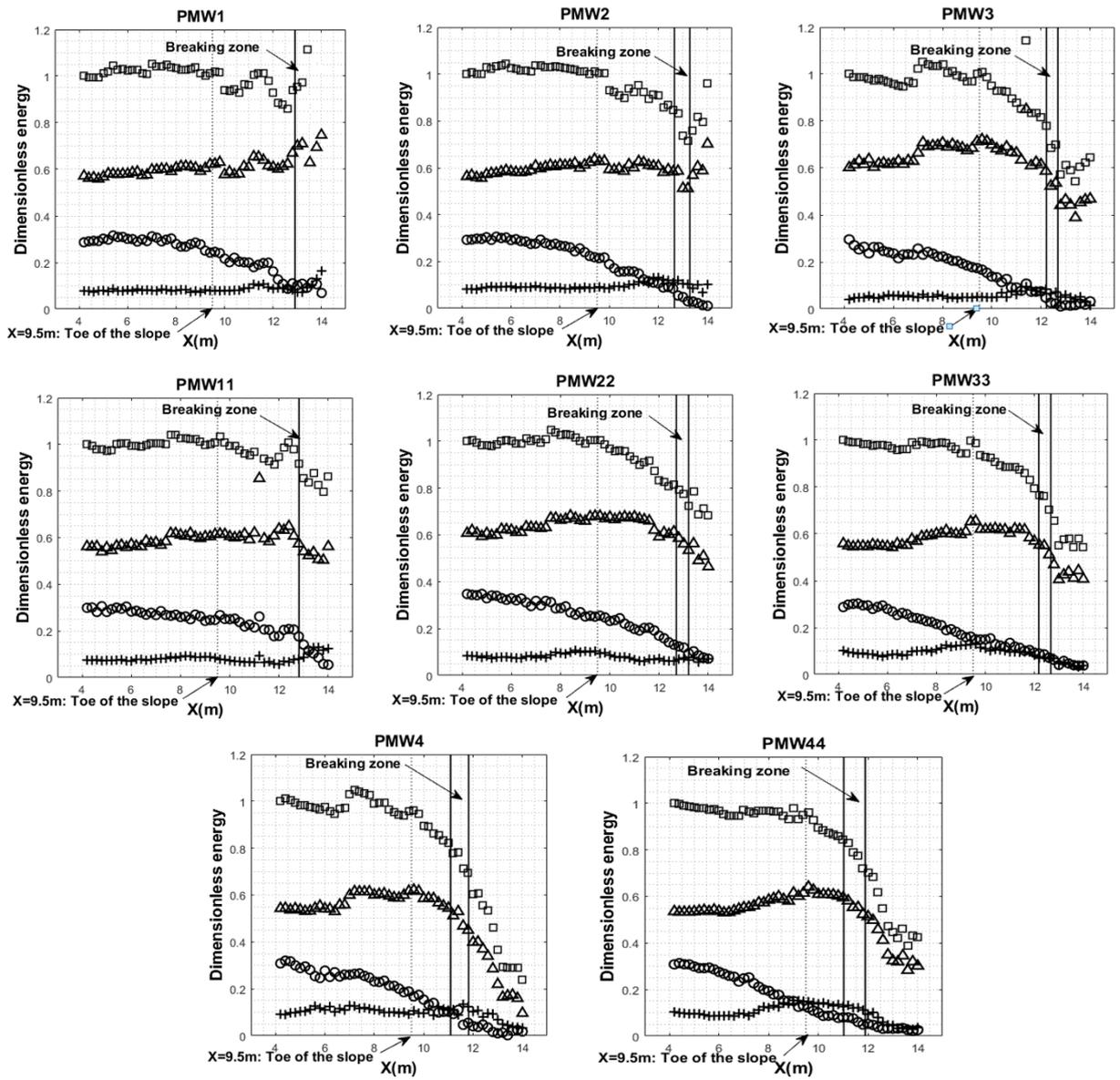

**Fig. 6**

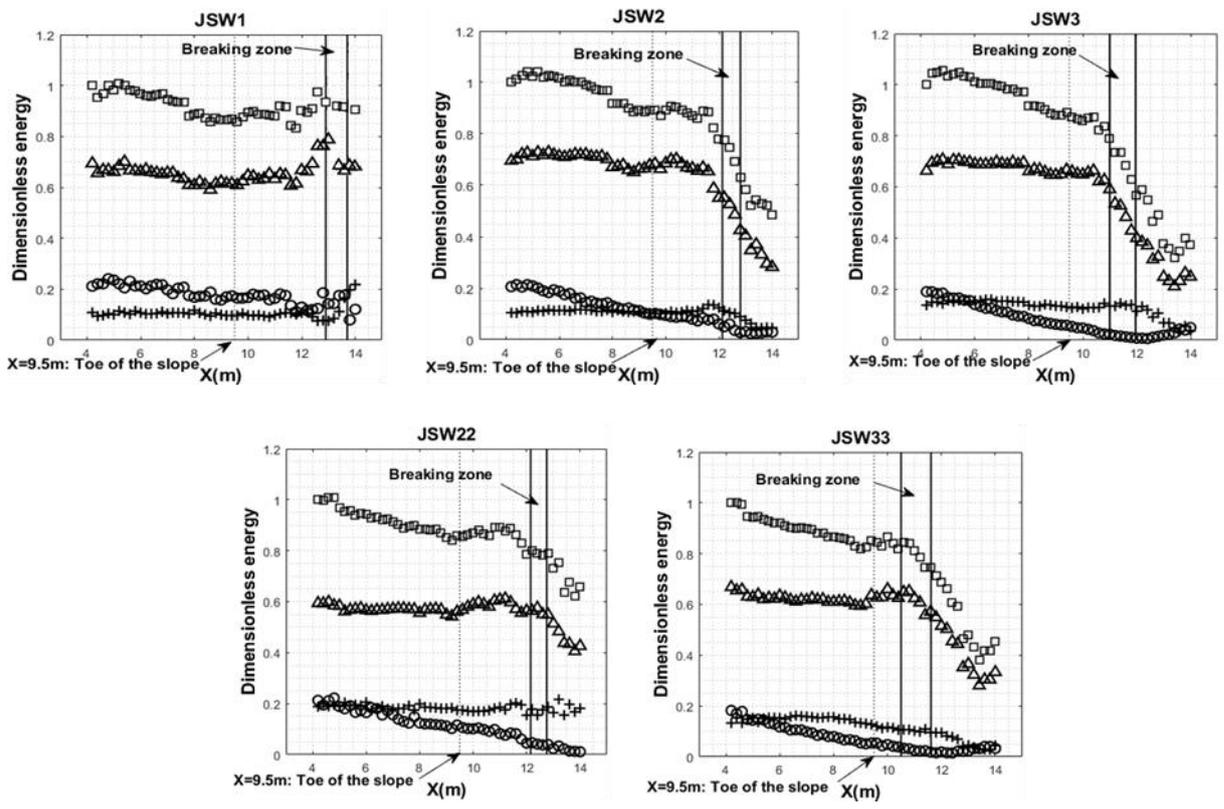

**Fig. 7**

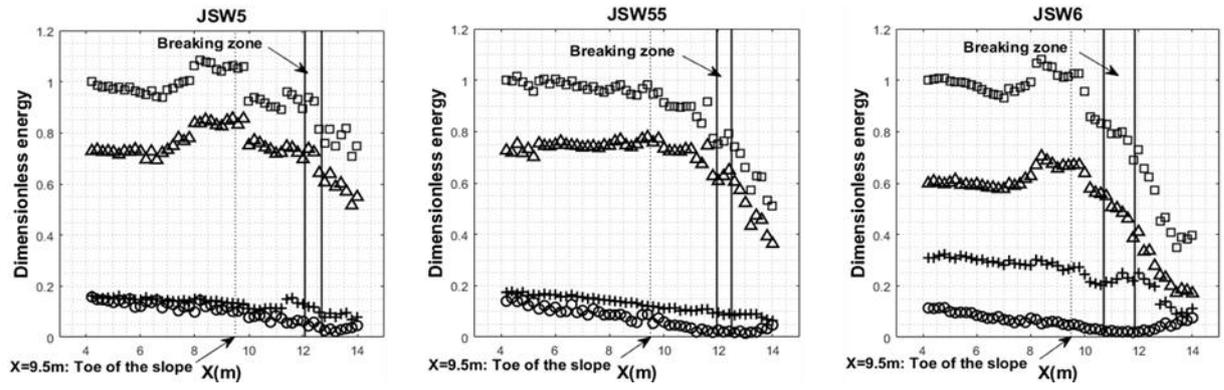

**Fig. 8**

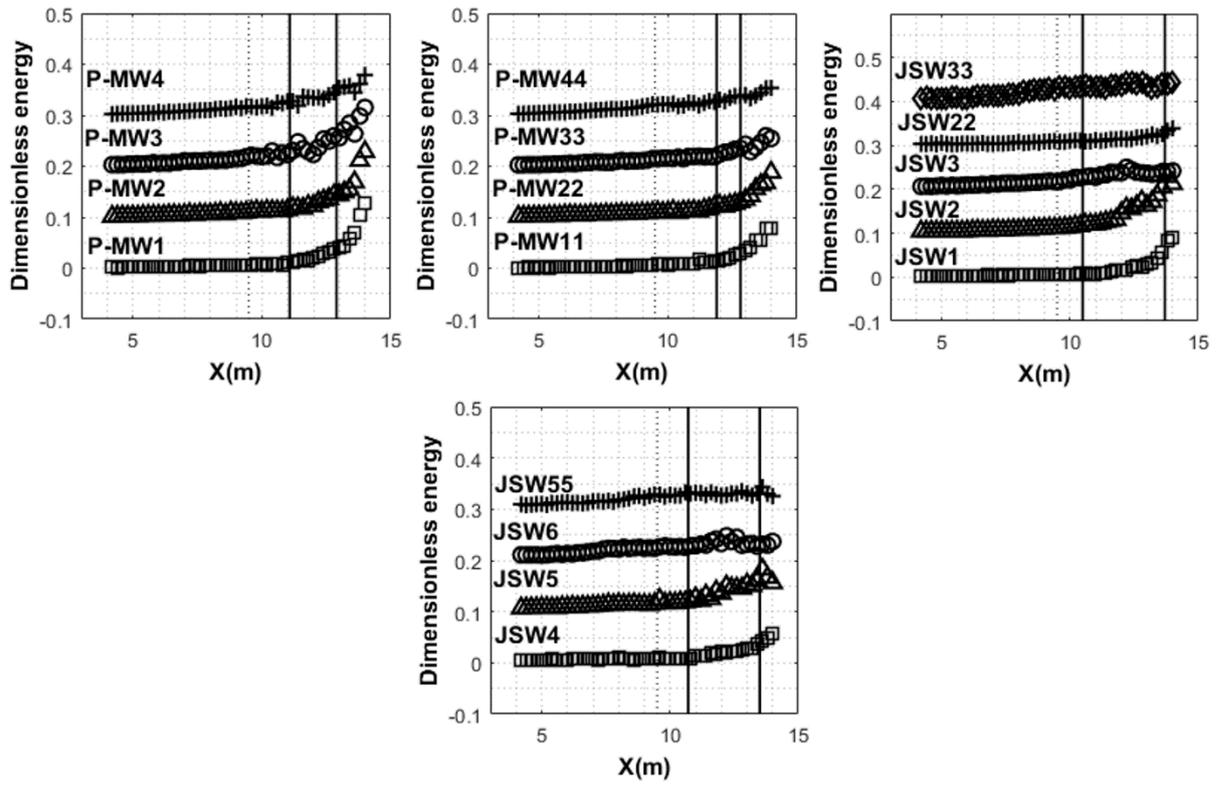

**Fig. 9**

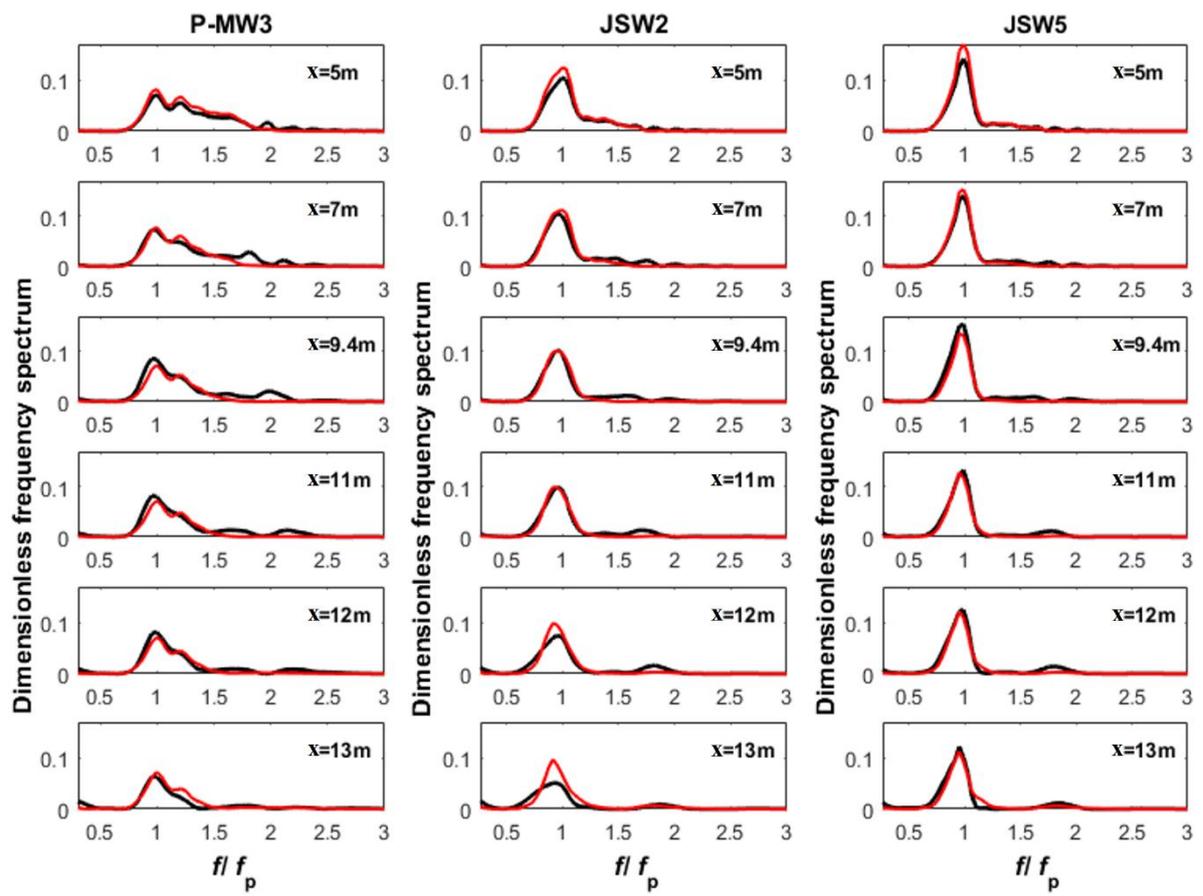

**Fig. 10**

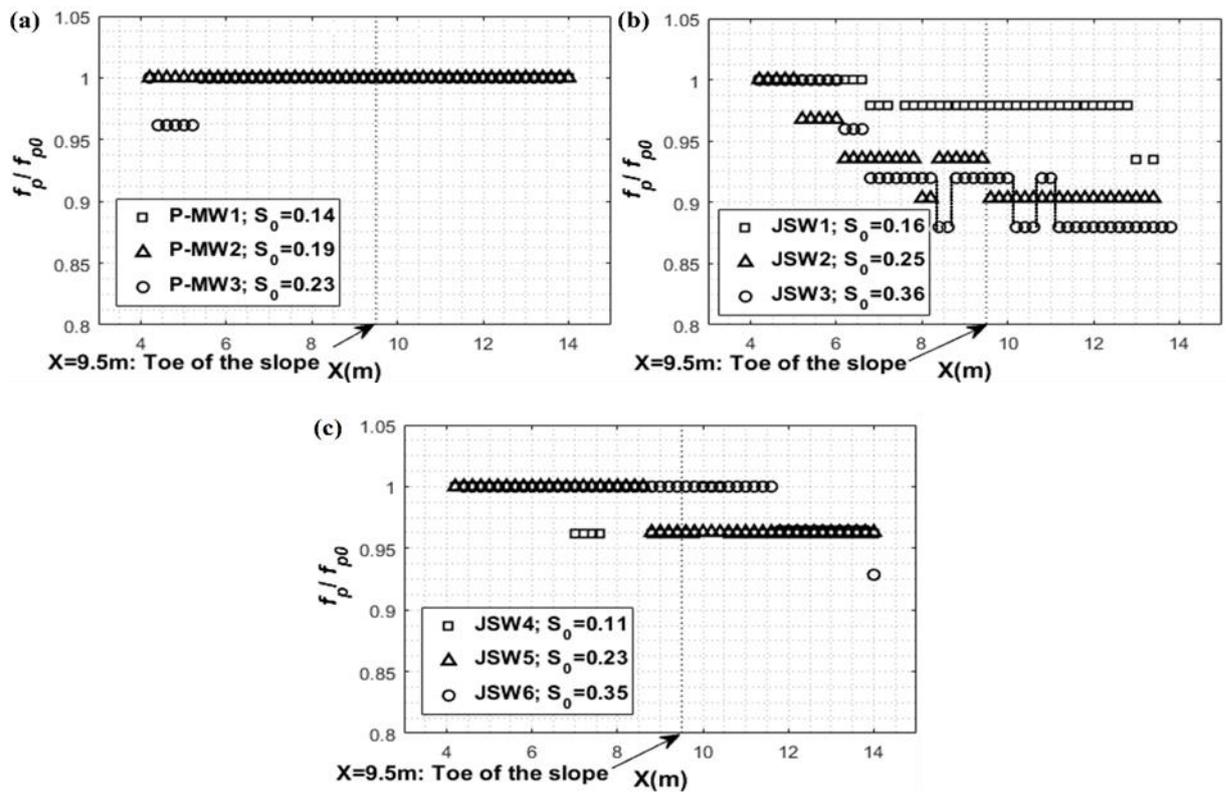

**Fig. 11**

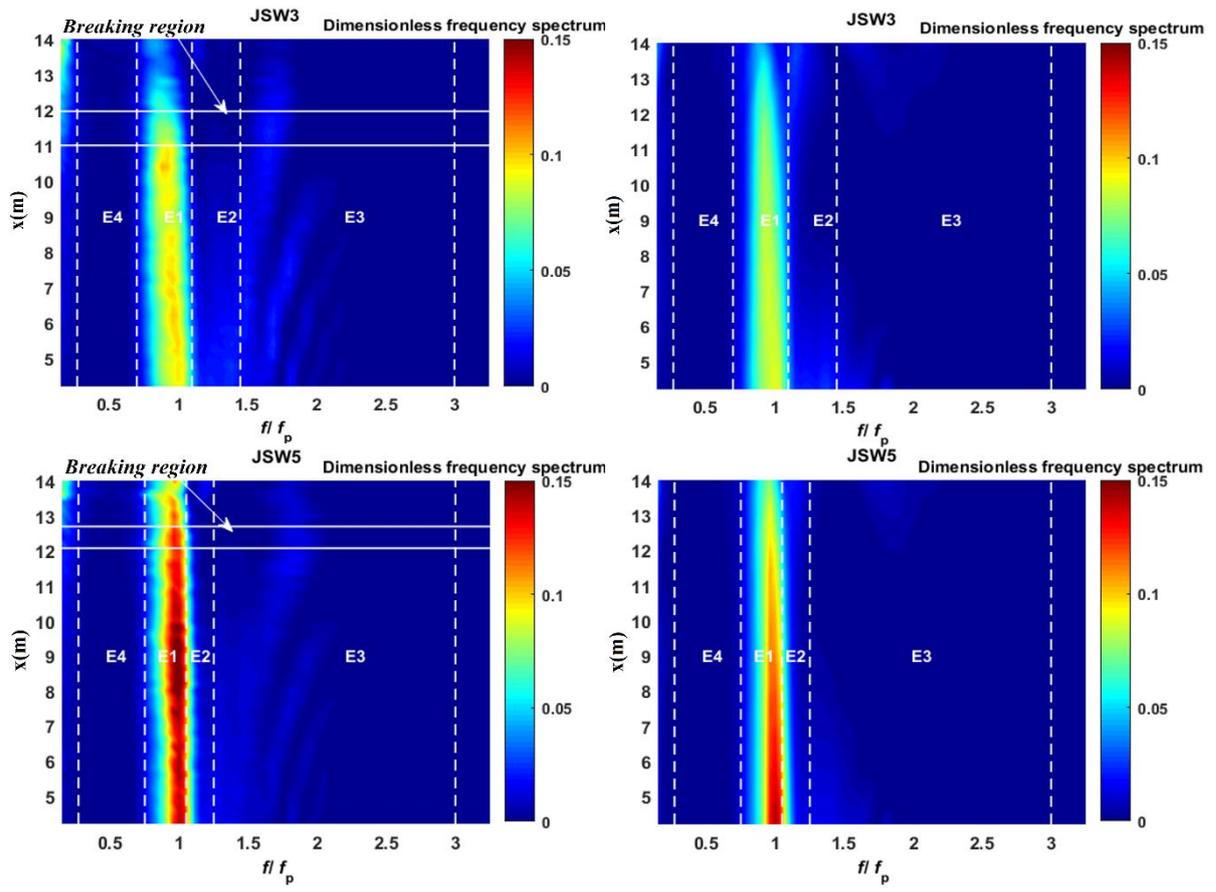

**Fig. 12**

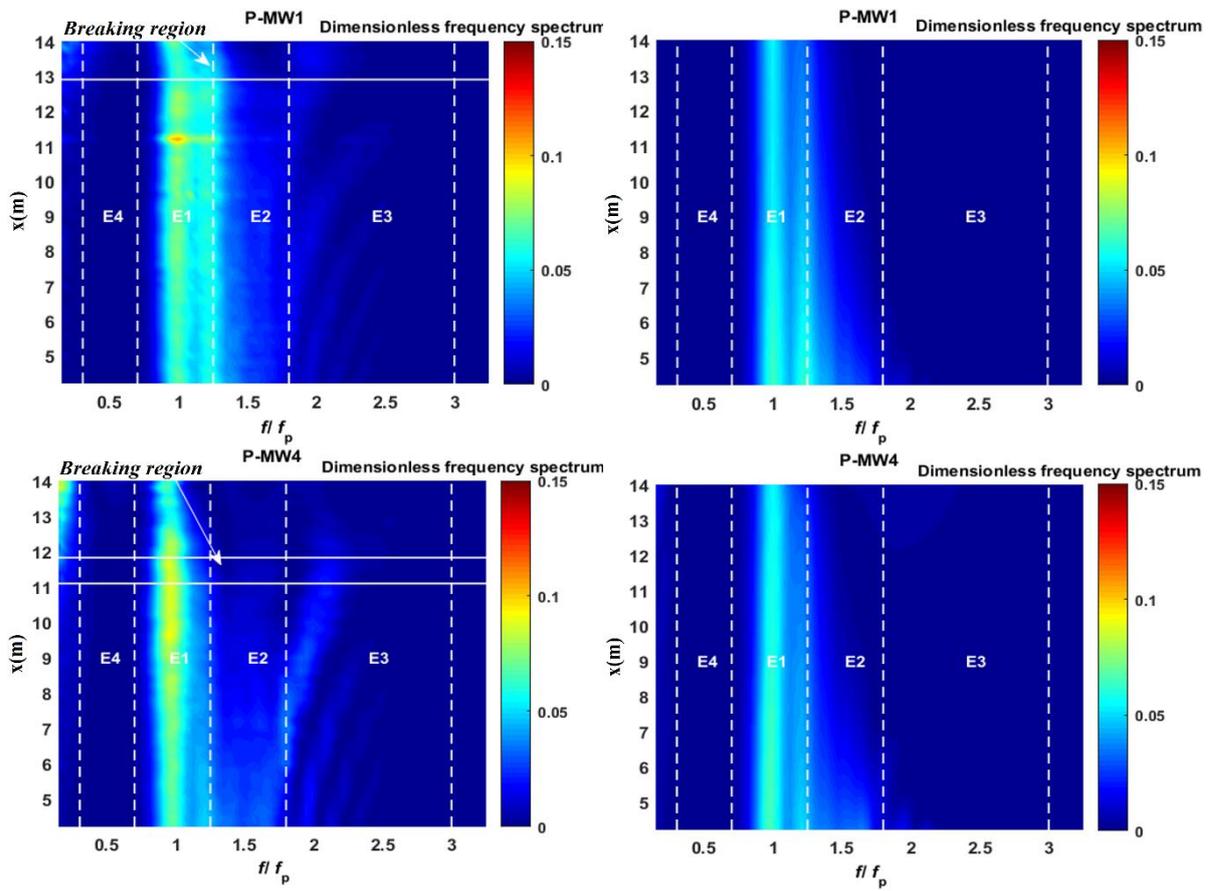

**Fig. 13**

| Wave train | Spectrum | $S_0$ | $f_p$ (Hz) | φ(rad) | $x_b$: Breaking locations (m) |
|---|---|---|---|---|---|
| P-MW1 | Pierson-Moskowitz | 0.14 | 0.66 | 0 | 12.9 |
| P-MW2 | Pierson-Moskowitz | 0.19 | 0.66 | 0 | [12.65..13.28] |
| P-MW3 | Pierson-Moskowitz | 0.23 | 0.66 | 0 | [12.22..12.69] |
| P-MW4 | Pierson-Moskowitz | 0.28 | 0.66 | 0 | [11.09..11.82] |
| P-MW11 | Pierson-Moskowitz | 0.14 | 0.66 | π | 12.81 |
| P-MW22 | Pierson-Moskowitz | 0.19 | 0.66 | π | [12.7..13.2] |
| P-MW33 | Pierson-Moskowitz | 0.23 | 0.66 | π | [12.18..12.67] |
| P-MW44 | Pierson-Moskowitz | 0.29 | 0.66 | π | [11.02..11.89] |
| JSW1 | JONSWAP (γ=3.3) | 0.16 | 0.75 | 0 | [12.9..13.7] |
| JSW2 | JONSWAP(γ=3.3) | 0.25 | 0.75 | 0 | [12.13..12.81] |
| JSW22 | JONSWAP(γ=3.3) | 0.26 | 0.75 | π | [12.07..12.76] |
| JSW3 | JONSWAP(γ=3.3) | 0.36 | 0.75 | 0 | [11..11.96] |
| JSW33 | JONSWAP(γ=3) | 0.38 | 0.75 | π | [10.5..11.61] |
| JSW4 | JONSWAP(γ=7) | 0.11 | 0.75 | 0 | 13.5 |
| JSW5 | JONSWAP(γ=7) | 0.23 | 0.75 | 0 | [12.07..12.69] |
| JSW6 | JONSWAP(γ=7) | 0.35 | 0.75 | 0 | [10.71..11.86] |
| JSW55 | JONSWAP(γ=7) | 0.27 | 0.75 | π | [11.95..12.49] |

Table 1